\documentclass[prl,twocolumn,showpacs,superscriptaddress]{revtex4}
\usepackage{graphicx}
\begin{document}

\title{Quantum Projection in an Ising Spin Liquid}
\date{\today}
\author{D.M. Silevitch}
\affiliation{The James Franck Institute and Department of Physics, The University of Chicago, 
Chicago, IL 60637}
\author{C. M. S. Gannarelli}
\author{A. J. Fisher}
\author{G. Aeppli}
\affiliation{London Centre for Nanotechnology and Department of Physics and Astronomy, UCL, 
London, WC1E 6BT, UK}
\author{T.F. Rosenbaum}
\email{tfr@uchicago.edu}
\affiliation{The James Franck Institute and Department of Physics, The University of Chicago, 
Chicago, IL 60637}

\begin{abstract}A transverse magnetic field is used to scan the diagonal and off-diagonal susceptibility of the uniaxial quantum magnet, $\text{LiHo}_{0.045}\text{Y}_{0.955}\text{F}_4$. Clusters of strongly-coupled spins act as the primary source for the response functions, which result from a field-induced quantum projection of the system into a classically forbidden (meaning non-Ising) regime. Calculations based on spin pairs reproduce only some features of the data and fail to predict the measured off-diagonal response, providing evidence of a multi-spin collective state. \end{abstract}

\maketitle

A basic feature of quantum mechanics is the superposition of states and the projection of these superpositions into observable quantities. From the Stern-Gerlach experiment of 1922, where quantum projection was first demonstrated with angular momentum states in silver atoms\cite{Gerlach22}, to spin echoes in nuclear magnetic resonance\cite{Hahn50}, where a series of radio frequency pulses induce the precession of nuclear spins projected into classically inaccessible states, strict experimental protocols have been derived for quantum manipulation at the atomic level. For quantum information purposes, it would be advantageous to extend these techniques to coherent clusters of atoms and to explore such effects in solids. In this paper, we pioneer measurements of the off-diagonal susceptibility to observe quantum mixing effects in a uniaxial magnetic salt, where we are able to rotate clusters of spins into a classically-forbidden (meaning non-Ising) state, transverse to the spin axis. The spins can be manipulated by a combination of ac and dc magnetic fields, detected by a multi-axis susceptometer, and parsed for their quantum-mechanical (xy) or classical (polarized along z, the Ising axis) nature by sweeping transverse magnetic field or temperature, respectively. Our technique takes full advantage of the tensor nature of the magnetic susceptibility, revealing the fundamental role played by the off-diagonal terms\cite{Ghosh03} in the Hamiltonian. Comparison of the results with calculations including only single ions and pairs of ions shows that the off-diagonal response is not only large, but that it also displays the characteristics of a quantum spin liquid\cite{Ghosh03}.

The results of the technique depend crucially on two properties of our system: (i) the ability to introduce quantum mechanics via a controllable, external tuning parameter, and (ii) the predilection of disorder to form spin clusters that may occur rarely but can dominate the physics. In particular, the random distribution of magnetic Ho ions in the diluted, spin-1/2 Ising system $\text{LiHo}_{0.045}\text{Y}_{0.955}\text{F}_4$ gives rise to regions where the local density of Ho is comparable to that of the pure compound, and hence are locally ferromagnetic or antiferromagnetic depending on the displacement between nearest neighbours. The Ising spins are constrained to point up or down along the crystallographic c-axis, and can be controllably mixed with the application of a magnetic field $H_t$ transverse to the c- (Ising) axis. The symmetry of the ground state doublet is then broken, not via a simple classical change in the relative occupancy of states as would result from the application of a longitudinal field, but via the creation of a ground state which is a quantum-mechanical admixture of up and down spins. Under proper initial conditions, it is then possible to project the net magnetic response of the system into the plane transverse to the Ising axis.

The parent compound, $\text{LiHoF}_4$, is a dipolar-coupled Ising ferromagnet with a Curie temperature $T_C = 1.53\:\text{K}$\cite{Hansen75}. The full Hamiltonian for the system includes crystal field, hyperfine, and dipolar and exchange interactions between the moments on the $\text{Ho}^{3+}$ ions. In the limit where the hyperfine and off-diagonal dipolar  interactions are ignored, the effective Hamiltonian for the system is then $H=-\sum_{i,j}J_{ij}\sigma_i^z\sigma_j^z-\Gamma\sum_i\sigma_i^x$, where $J_{ij}$ is a dipole-dipole interaction operator and $\Gamma$ is the magnetic field-induced splitting energy of the ground-state doublet. For $H_t < 10\:\text{kOe}$, $\Gamma\propto H_T^2$\cite{Wu91}.

The magnetic ground state of $\text{LiHoF}_4$ is extremely sensitive to substitution of non-magnetic Y for the magnetic Ho ions, moving successively from disordered ferromagnet to spin glass to spin liquid with decreasing x\cite{Reich90}. We focus here on the spin-liquid (antiglass) in $\text{LiHo}_{0.045}\text{Y}_{0.955}\text{F}_4$, and map out the spin dynamics using a combination of non-linear magnetic susceptibility along the Ising axis and linear measurements of multiple components of the susceptibility tensor.  The full vector magnetization can be written as an expansion of susceptibilities about H=0, including non-linear contributions:
\begin{eqnarray}
M_l &=& \chi_l^{(1,0)}h_l + \chi_l^{(3,0)}h_l^3 +  \chi_l^{(1,2)}h_lh_t^2+\cdots\nonumber\\
M_t &=&\chi_t^{(0,1)}h_t + \chi_t^{(0,3)}h_t^3 + \chi_t^{(2,1)}h_l^2h_t +\cdots\label{eq:vector}
\end{eqnarray}
In this experiment, the ac probe fields are always applied along the longitudinal (Ising) axis, but we probe the full tensor susceptibility, $\chi_{ij}=\partial M_i/\partial h_j$  at a finite static field.  The off-diagonal element $\chi_t\equiv\partial M_t/\partial h_l$ of this finite-field susceptibility has leading term $2\chi_t^{(2,1)}h_lh_t$. 

\begin{figure}
\includegraphics[scale=0.85]{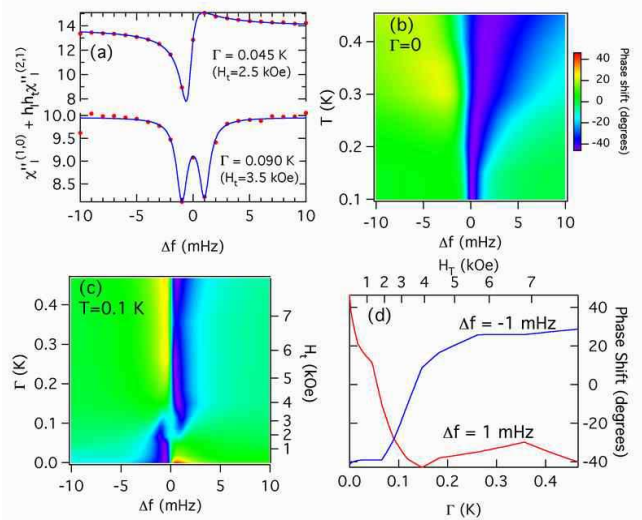}
\caption{\label{fig:pumpprobe}(Color online) Nonlinear magnetic response parallel to the Ising axis in a pumped spin liquid. (a) Dissipative component of the longitudinal magnetic response $\chi_l=\partial M_l/\partial h_l$  in the presence of a 0.45 Oe 202 Hz pump field in the longitudinal (Ising) direction; $\Delta f$ is the frequency difference between the pump and probe. The integration time for each point ranged from 2000 to 8000 seconds, which needs to be included in detailed analysis of the data within a few mHz of the pump frequency. Curves: fits to Fano resonance forms. (b) Phase shift of the nonlinear response as a function of frequency and temperature for transverse field $\Gamma=0$. The linewidth increases exponentially with increasing T. (c) Phase shift of the nonlinear response as a function of frequency and $\Gamma$ at fixed T=100 mK. Here the linewidth remains unchanged and the resonance maintains coherence. An apparent level crossing occurs at $\Gamma\sim 0.1\:\text{K}$ ($H_t\sim 4\: \text{kOe}$). (d)  Phase shift of the nonlinear response at $\Delta f=\pm1\:\text{mHz}$ as a function of $\Gamma$.}
\end{figure}

Previous studies of the dynamics\cite{Reich90,Ghosh02} have observed a continuum of excitations corresponding to a set of individual oscillators spread throughout the crystal. That this is an oscillator continuum rather than a relaxation spectrum was derived from hole burning, a technique commonly applied in optical spectroscopy, but adapted for audio frequency characterization of magnetic systems\cite{Ghosh02}. The frequency-dependent susceptibility, measured in the presence of an ac pump, is exemplified in Fig.~\ref{fig:pumpprobe}(a), and generally can be described by a single Fano resonance\cite{Fano61}, corresponding to an oscillator, at the pump frequency, coupled to the bath of the remaining oscillators. With increasing temperature (Fig.~\ref{fig:pumpprobe}(b)), the linewidth of the resonance increases exponentially with temperature and coherence is lost. By contrast, with increasing transverse field at low temperature (Fig.~\ref{fig:pumpprobe}(c)) the linewidth remains constant over the entire range: the mode remains well-defined and coherent even at effective energies where temperature has washed out the resonance.  Moreover, there is a phase shift in the nonlinear response at a well-defined crossover field of $H_t\sim3.5\:\text{kOe}$ (Fig.~\ref{fig:pumpprobe}(d)), suggesting a quantum-mechanical level crossing at a discrete value of $H_t$.

The non-linear response which undergoes a change at a discrete value of the transverse field underscores the fundamental quantum nature of the system and identifies the temperatures and transverse fields of interest.  We now can project the collective modes into the plane perpendicular to the Ising axis and follow two orthogonal components of the ac magnetic susceptibility tensor simultaneously by using a multi-axis susceptometer. This unique instrument uses a solenoid to provide an ac probe field along the Ising axis, and a pair of nested pickup coils to measure both the diagonal response along the Ising axis ($\chi_l\equiv\partial M_l/\partial h_l=\chi_l^{(1,0)}+3h_l^2\chi_l^{(3,0)}+\cdots$) and the off-diagonal response in the transverse plane ($\chi_t\equiv\partial M_t/\partial h_l=2\chi_t^{(2,1)}h_lh_t+\cdots$). This off-diagonal response is measured parallel to the large dc transverse magnetic field $H_t$.

\begin{figure}
\includegraphics[scale=0.4]{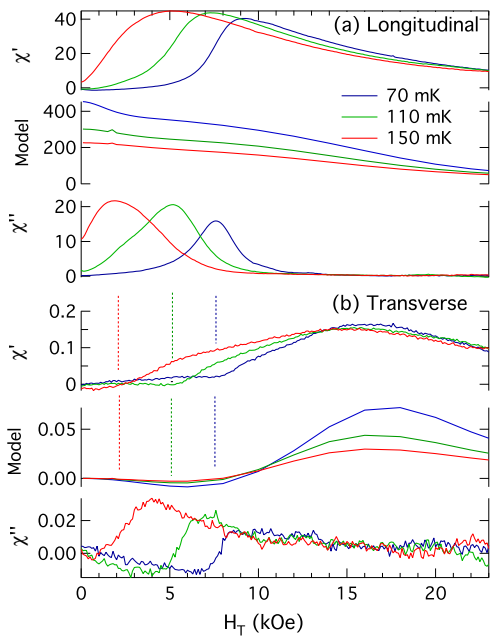}
\caption{\label{fig:vector}(Color online) Measured and simulated magnetic response vs. transverse magnetic field at three temperatures with a polarizing dc longitudinal field (a) Response along longitudinal (Ising) axis, measuring the diagonal component of the susceptibility tensor ($\chi_l=\partial M_l/\partial h_l$). (b) Response in transverse plane, measuring the off-diagonal element $h_lh_t\chi_t^{(2,1)}$. The imaginary part of the transverse susceptibility tracks the real part of the longitudinal susceptibility, demonstrating that approximately 1\%
of the spins coherently rotate $90^\circ$ in space, $90^\circ$ out of phase. For both (a) and (b), the model curves are a 2-ion nearest-neighbor calculation as described in the text. }
\end{figure}

We plot in Fig.~\ref{fig:vector} the full complex tensor components  $\chi_l$ and $\chi_t$   measured at constant temperature as a function of dc magnetic field. The magnetic field was tilted at an angle of $0.6^\circ$ relative to the transverse plane of the sample, corresponding to a field along the Ising axis equal to 1\%
 of the field in the transverse direction. The shape of $\chi_l$ (Fig. 2a) is itself remarkable. Instead of simply a speed-up of the dynamics due to quantum tunneling between the spin up and spin down (along z) eigenstates, we see what looks like a spectrum (as a function of $H_t$), with a gap-like threshold that becomes progressively better defined on cooling. The imaginary response, $\chi^{\prime\prime}_l$, peaks at the midpoint of the leading edge of the real component, $\chi^\prime_l$. 

The behavior of the transverse signal, however, is significantly different and quantum-mechanical in the sense that a classical Ising system should not exhibit any transverse response to an excitation field in the longitudinal direction. The real component of the transverse response (Fig.~\ref{fig:vector}b) has a sharp onset at a well-defined magnetic field, apparently set by the level crossing shown in Fig~\ref{fig:pumpprobe} and coincident with the threshold field in $\chi^\prime_l$. The imaginary component of the transverse susceptibility, $\chi^{\prime\prime}_t$, strikingly has features echoing those in the real part of the longitudinal susceptibility, $\chi^\prime_l$, displaying a zero crossing at the threshold field.  Both are characterized by larger transverse fields at lower temperatures, peaking at the same values of $H_t$, but reduced in magnitude by a hundred fold. One per cent of the spins appear to rotate $90^\circ$ in space while at the same time rotating $90^\circ$ in phase.

\begin{figure}
\includegraphics[scale=0.38]{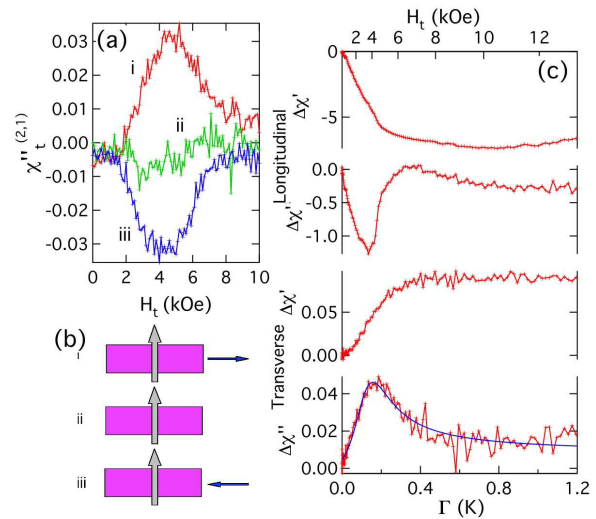}
\caption{\label{fig:phase}(Color online) Manipulation of the quantum mechanically projected spin clusters. (a) $\chi_t^{\prime\prime}$  as a function of transverse field at T = 150 mK with a polarizing field in the longitudinal (Ising) direction of (i) +50 Oe, (ii) 0 and (iii) -50 Oe. The projected susceptibility flips $180^\circ$ when the polarizing field flips $180^\circ$. (b) Schematic showing the field orientations. The vertical arrows show the direction of the transverse field and the horizontal arrows the longitudinal field. (c) $\Delta\chi$, the difference between the susceptibility in geometry (ii) and geometry (i), plotted versus the doublet energy splitting $\Gamma$; $\Gamma\propto H_t^2$ for small $H_t$. The curve is a fit to a Fano resonance, which indicates a marked decrease in the coupling to the spin bath at the lowest temperature (see text for details).}
\end{figure}

Inspection of equation \ref{eq:vector} shows that reversing the direction of the dc longitudinal bias field by $180^\circ$ should shift the phase of the transverse response by $180^\circ$.  We see in Fig.~\ref{fig:phase}(a) exactly such a relationship.  With zero longitudinal field, there is no net moment---the number of spin up clusters balance the number of spin down clusters---and there is no transverse signal. Adding a field of 50 Oe along the longitudinal axis produces a strong signal in the transverse plane; changing the polarity of this field reverses the polarity of the transverse susceptibility. This response scales with $h_l$, eventually saturating at $|h_l|=100\:\text{Oe}$. Our picture then is that the transverse signal is due to the polarization of a fraction of the spins by the longitudinal field, followed by tilting into the plane by the dc transverse field. As the small longitudinal ac drive field alternately raises and lowers the projection of the net moment onto the z-axis, the net moment will move along an arc with a projection onto the transverse plane, and it is this which gives rise to the off-diagonal response measured via $\chi^\prime_t$.

This behavior contrasts markedly with the behavior of isolated $\text{Ho}^{3+}$ ions. NMR and SQUID magnetometry of $\text{LiHo}_{0.002}\text{Y}_{0.998}\text{F}_4$ found strong resonances at a set of longitudinal fields corresponding to nuclear hyperfine level crossings in the single-ion energy diagram\cite{Giraud01,Graf06}. These level crossings occur at integer multiples of 230 Oe, substantially larger than the 50-80 Oe at which the effects discussed here emerge, and are discrete rather than continuous. Hence, the observed effects do not appear to reflect the intrinsic properties of the $\text{Ho}^{3+}$ ion, but instead represent a collective phenomenon.

In order to see this quantitatively, we calculated the diagonal and off-diagonal response, as shown in Fig.~\ref{fig:vector}. This calculation involves an exact numerical diagonalization of the full microscopic Hamiltonian, incorporating the intrinsic susceptibility of individual $\text{Ho}^{3+}$ ions and nearest-neighbor pairs (averaged over nearest-neighbor vectors).  Antiferromagmetically coupled pairs exhibit a diagonal response comparable to what is experimentally observed; however, such pairs do not dominate the average and so the pair model fails to capture the suppression of the susceptibility at low transverse fields.  In such a pairs-only model the true transverse susceptibility vanishes due to inversion symmetry; however, there is a contribution from $\chi_{xx}$, sampled through a small tilt in the probe field. This contribution is shown in Fig.~\ref{fig:vector}(b) for a field tilted by $0.6^\circ$ in the transverse direction. The computed result is considerably smaller than observed and does not properly reproduce the sharp onset in $\chi$, suggesting that the transverse response is dominated by larger clusters of spins. However the peak in the transverse response occurs at a similar field scale to that in the measurements. This characteristic transverse field is required to break up the antiferromagnetic ordering of a pair in the basal plane (i.e. along $[100]$). These discrepancies suggest that the actual behavior is dominated by a strongly correlated (entangled) state involving multiple spins, consistent with the zero-field dynamics in the same material\cite{Ghosh02}. The ability to see evidence of this entangled state in the off-diagonal response suggests using this technique as a spectroscopic probe of the spin-liquid state.

The direct link between the longitudinal and transverse responses can be illuminated by plotting the difference in susceptibility $\Delta\chi$ between $h_l=0$ and $h_l=50\:\text{Oe}$ (Fig.~\ref{fig:phase}(c)). The 50 Oe longitudinal field acts to decrease $\chi_l$, accompanied by a proportionate 1\%
 gain in $\chi_t$. This proportionality provides a direct measurement of the diagonal non-linear susceptibility, $\chi^{(3,0)}_l$, and indicates that the transverse signal arises from a non-linear excitation initially present along the Ising axis and subsequently rotated into the transverse plane by the external magnetic fields. 

Additional insight into the nature of the transverse excitations can be found by looking at the imaginary component, shown in the bottom panel of Fig.~\ref{fig:phase}(c). The observed lineshape can be accurately fit by the dissipative component of a Fano resonance\cite{Fano61}, 
\begin{equation}
\chi^{\prime\prime}_T\propto\left(\frac{q\Gamma_r}{2}+(\Gamma-E_r)\right)^2
\left/\left( (\Gamma-E_r)^2 + \left(\frac{\Gamma_r}{2}\right)^2\right)\right. .
\end{equation}
$E_r$, the energy of the resonance, depends strongly on T, taking values that correspond to the location of the onset of the transverse $\chi^\prime$. As with the pump-probe measurements shown in Fig~\ref{fig:pumpprobe}, the Fano lineshape in the off-diagonal response suggests that the underlying physical mechanism again consists of a resonance due to a set of tightly-bound clusters, and coupling between the resonance and the continuum of excitations in the spin bath Ð the physical origin of which is most likely the nuclear spin degrees of freedom\cite{Prokofiev00,Ronnow05}. The inherent linewidth of the resonance, $\Gamma_r$, was found to be approximately 0.25 K, with no significant variation due to T. By comparison, $q$, the coupling between the bath and the resonance is 1.3 at T=0.15 K and T=0.11 K, falling to 0.8 for T = 0.07 K.

\begin{figure}
\includegraphics[scale=0.4]{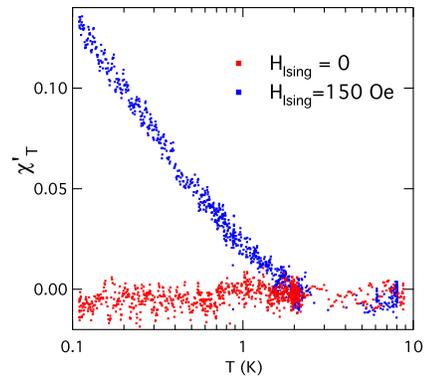}
\caption{\label{fig:temp}(Color online) Onset of a transverse susceptibility. $\chi_t^\prime$, shown here for $H_t = 15$ kOe, becomes appreciable for $T < 1.5$ K, where the interaction energy with the external field becomes comparable with the thermal energy scale. In the absence of a longitudinal field, $\chi_t^\prime$ vanishes.}
\end{figure}

The susceptibility from rotations in the transverse plane represents approximately 1\%
 of the total spectral intensity. This is of the same order as the spectral intensity lost in the same material when it is pumped by an oscillatory magnetic field large enough to induce hole-burning\cite{Ghosh02}. In order to test further whether the same coherent spin clusters are contributing to the non-linear diagonal and the linear off-diagonal susceptibilites, we have studied the temperature dependence of the transverse susceptibility. As shown in Fig.~\ref{fig:temp}, the transverse response has an onset temperature of approximately 1.5 K, $T_C$ of pure $\text{LiHoF}_4$. This onset temperature corresponds to the point where the polarization energy calculated for a single ion interacting with the external field matches the thermal energy scale. 

We have shown that a combination of longitudinal and transverse magnetic fields can be used to create excitations of spin clusters in $\text{LiHo}_x\text{Y}_{1-x}\text{F}_4$. Their temporal evolution can be controlled\cite{Ghosh02} by direct analogy to spin echo experiments in nuclear magnetic resonance\cite{Slichter96}. In a Heisenberg spin system, application of a transverse magnetic field results in the spins continuously rotating towards the field direction, and the linear response has been used to characterize a wide variety of spin and charge effects in liquids and in spin ladders\cite{Takigawa98}. In an Ising spin system, a linear response away from the longitudinal axis is forbidden, but higher order terms in the susceptibility can be detected transverse to the Ising axis following specific protocols. Accessing this off-diagonal magnetic response and combining it with the addressable nature of the resulting excitations\cite{Ghosh03} suggests the possibility of selective creation and projection of mixed states, a proven strategy for manipulating quantum information\cite{Gershenfeld97}. 

\begin{acknowledgments}The work at the University of Chicago was supported by U.S. DOE Grant No. DE-FG02-99ER45789, and that at UCL by EPSRC grant EP/D049717/1.\end{acknowledgments}

\end{document}